\input{aipcheck}
\documentclass[
    ,final            % use final for the camera ready runs
  ]
  {aipproc}

\layoutstyle{6x9}

\begin{document}

\title{The International X-ray Observatory}

\classification{95.55.Ka, 95.85.Nv}
\keywords      {Astronomical and space-research instrumentation, X-ray telescopes }

\author{Nicholas E. White}{
  address={NASA Goddard Space Flight Center, Greenbelt, MD, USA}
}

\author{Arvind Parmar}{
  address={ESAC - PO Box 78, E-28691 Villanueva de la Canada, Madrid, Spain}
}

\author{Hideyo Kunieda}{
  address={X-ray Astronomy Group, Nagoya University, Nagoya, Japan}
}

\author{Kirpal Nandra}{
  address={Astrophysics Group, Imperial College London, London, UK}
}

\author{Takaya Ohashi}{
  address={Physics Department, Tokyo Metropolitan University, Tokyo, Japan}
}

\author{Jay Bookbinder}{
  address={Smithsonian Astrophysical Observatory, Cambridge, MA, USA}
}

%\author{<author3>}{
  %address={<common address for author2 and author3>}
%  ,altaddress={<author1 address>} % additional visiting address
%}

\begin{abstract}
The International X-ray Observatory (IXO) is a joint ESA-JAXA-NASA effort to address fundamental and timely questions in astrophysics: What happens close to a black hole? How did supermassive black holes grow? How does large scale structure form? What is the connection between these processes? To address these questions IXO will %: trace orbits close to the event horizon of black holes, measure black hole spin for several hundred active galactic nuclei (AGN), use spectroscopy to characterize outflows and the environment of AGN during their peak activity, search for super-massive black holes out to redshift $z = 10$, map bulk motions and turbulence in galaxy clusters, find the missing baryons in the cosmic web using background quasars, and observe the process of cosmic feedback where black holes inject energy on galactic and intergalactic scales. IXO will 
employ optics with 3 sq m collecting area and 5 arc sec angular resolution - 20 times more collecting area at 1 keV than any previous X-ray observatory. Focal plane instruments will deliver a 100-fold increase in effective area for high-resolution spectroscopy, deep spectral imaging over a wide field of view, unprecedented polarimetric sensitivity, microsecond spectroscopic timing, and high count rate capability. The mission is being planned for launch in 2021 to an L2 orbit, with a five-year lifetime and consumables for 10 years. %Previous experience assures us that unexpected discoveries will abound -- a key feature of great observatories.
\end{abstract}

\maketitle

%%%%%%%%%%%%%%%%%%%%%%%%%%%%%%%%%%%%%%%%%%%%
%% MAINMATTER
%%%%%%%%%%%%%%%%%%%%%%%%%%%%%%%%%%%%%%%%%%%%

\section{Science Objectives}

The driving science goals of IXO are to determine the properties of the extreme environment and evolution of black holes, measure the energetics and dynamics of the hot gas in large cosmic structures, and establish the connection between these two phenomena. In addition, IXO measurements of virtually every class of astronomical object will return serendipitous discoveries, characteristic of all major advances in 
%astronomical 
observing
capabilities. 

%\paragraph
{\it Matter Under Extreme Conditions}\ The observational consequences of strong gravity can be seen close to the event horizon, where the extreme effects of General Relativity (GR) are evident in the form of gravitational redshift, light bending, and frame dragging. 
%The spectral signatures needed to determine the physics of the accretion flow into the black hole are only found in X-rays. 
Observations of SMBH with XMM-Newton have revealed evidence of ``hot spots'' on the disk that light up in the iron K$\alpha$\ line, allowing us to infer their motions. Each parcel of gas follows a nearly circular orbit around a black hole. Tracing these on sub-orbital timescales, however, requires the large 0.65 m$^2$\ effective area at $\sim$ 6 keV provided by IXO. The emission from these hot spots appears as ``arcs'' in the time-energy plane. GR makes specific predictions for the form of these arcs, and the ensemble of arcs reveals the mass and spin of the black hole and the inclination of the accretion disk \citep{AR03}. Deviations from the GR predictions will create apparent changes in these parameters as a function of time or hot spot radius. %IXO will enable the first orbitally time-resolved studies of 10-20 SMBH and provide a direct probe of the physics of strong gravity.

%\begin{figure}
%\includegraphics[totalheight=2in]{WaterfallSim_big_v5.pdf}
%\caption{IXO will resolve multiple hot spots in energy and time as they orbit the SMBH, each of which traces the Kerr metric at a particular radius. In the time-energy plane, the emission from these hot spots appears as "arcs", each corresponding to an orbit of a given bright region.}
%\end{figure}

%\paragraph
{\it Neutron State Equation of State}\ Neutron stars have the highest known matter densities in nature, beyond the densities produced in terrestrial laboratories, but the physics are uncertain due to the complexity of Quantum Chromodynamics (QCD) in this regime, which leads to different equations of state. IXO will determine the mass-radius relationship for dozens of neutron stars of various masses with four distinct methods \citep{Paerels09}: (1) the gravitational redshift (2) Doppler shift and broadening of atmospheric absorption lines, (3) pulse timing distortions due to gravitational lensing, and (4) pressure broadening of line profiles. %An X-ray polarimeter on IXO will make observations of magnetars that also test predictions of QED. 
%Magnetars have magnetic fields $B > 4.4 \times 10^{13}$\,G where QED predicts novel effects, such as vacuum birefringence. In the presence of matter, resonant polarization mode conversion will occur that will be observed with the polarimeter on IXO.

%Some url test \url{http://www.world.universe}.

\begin{figure}
\includegraphics[totalheight=2in]{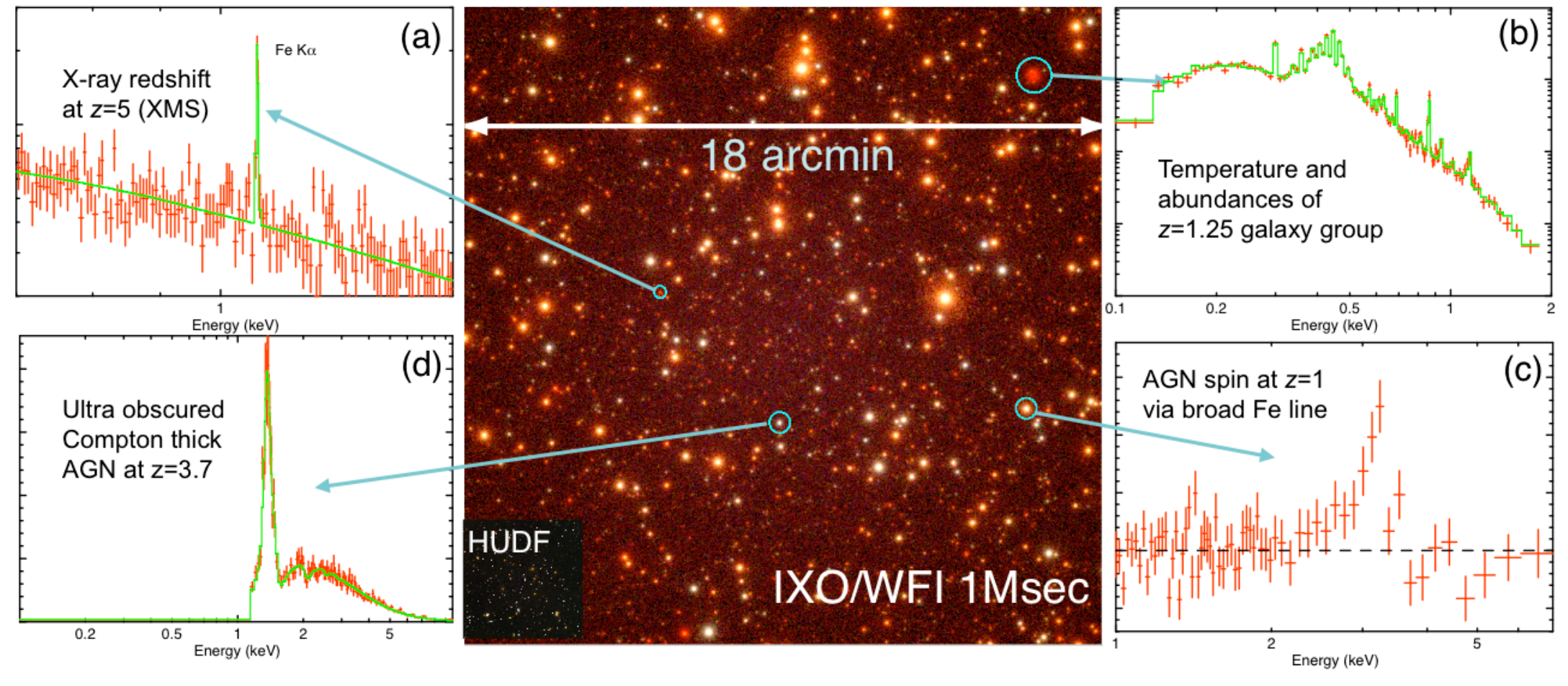}
\caption{WFI Simulation of the Chandra Deep Field South with Hubble Ultra Deep Field (HUDF) in inset. Simulated spectra of various sources are shown, illustrating IXO's ability (clockwise from top left) to: a) determine redshift autonomously in the X-ray band, b) determine temperatures and abundances even for low luminosity groups to $z>1$, c) make spin measurements of AGN to a similar redshift, and d) uncover the most heavily obscured, Compton-thick AGN.\label{fig:cdfs}}
\end{figure}

%\paragraph
{\it Black Hole Evolution}\ Luminous, $\sim10^9$\,M$_{\odot}$\  Super-Massive Black Holes (SMBHs) have been detected at $z\sim 6$. Growing such massive SMBHs within $<1$\,Gyr requires sustained Eddington-limited accretion.  IXO will chart the growth of Supermassive Black Holes over cosmic time \citep{Nandra09}. Finding growing SMBH at $z > 7$, which are rare objects, drives the combination of large effective area (3 m$^2$\ at 1 keV), good angular resolution (5~arc~sec) and large field of view (18 arcmin) specified for IXO. These capabilities allow IXO to reach Chandra's limiting sensitivity 20 times faster (Fig.~\ref{fig:cdfs}), enabling the first full characterization of the population of accreting SMBHs at $z\sim 7$, and constraints at $z = 8-10$, deep into the cosmic ``dark age.'' At lower redshifts, $z = 1-3$, where the majority of accretion and star formation in the Universe occurs, IXO's high throughput for imaging and spectroscopy will uncover and characterize the properties of the most obscured AGN, observing $\sim 10,000$\ AGN in a 1 Msec 1 sq. deg survey. 

A key observational signature is the iron K$\alpha$ emission line, produced via the illumination of the disk by the primary X-ray continuum and distorted in energy and strength by the gravitational field and relativistic motions around the black hole. The line profile can be used to measure the black hole spin. The $z\sim 0$\ spin distribution is a record of the relative importance of mergers versus accretion in the growth history of black holes. IXO will measure the black hole spin of $\sim 300$\ SMBHs \citep{Brenneman09}, sufficient to distinguish merger from accretion models. 

%\paragraph
{\it Cosmic Feedback from SMBHs}\ Energetic processes around black holes result in huge radiative and mechanical outputs \citep{Miller09}, which can potentially have a profound effect on their larger scale environment in galaxies, clusters and the intergalactic medium \citep{Fabian09}. The black hole can heat surrounding gas via its radiative output, and drive outflows via radiation pressure. Mechanical power emerging in winds or jets can also provide heating and pressure. The high spectral resolution and imaging of IXO will provide the spectral diagnostics needed to distinguish between them. 

For outflows that are radiatively accelerated in AGN, X-ray observations will determine the total column density and flow velocity, and hence the kinetic energy flux. IXO will be sensitive to ionization states from Fe I to Fe XXVI over a wide redshift range, allowing the first determination of how feedback affects all phases of interstellar and intergalactic gas, from million-degree collision-ionized plasmas to ten-thousand degree photo-ionized clouds. IXO measurements will probe over 10 decades in radial scale, from the inner accretion flow where the outflows are generated, to the halos of galaxies and clusters where the outflows deposit their energy. %In the centers of many galaxy clusters, the radiative cooling time of the X-ray-emitting gas is much shorter than the age of the system. Despite this, the gas there is still hot. Mechanical power from the central AGN acting through jets is thought to somehow compensate for the energy lost across scales of tens to hundreds of kpc. IXO will map the gas velocity across dozens of galaxy clusters to an accuracy of tens of km/s, revealing how the mechanical energy is spread and dissipated. 

%\paragraph
{\it Galaxy Cluster Evolution}\ 
Structure formation is a multi-scale problem. Galaxy formation depends on the physical and chemical properties of the intergalactic medium (IGM). The IGM in turn is affected by energy and metal outflows from galaxies. Detailed studies 
of the IGM in galaxy clusters are now limited to the relatively nearby Universe ($z < 0.5$). IXO will measure the thermodynamic properties and metal content of the first low-mass clusters emerging at $z \sim 2$ and directly trace their evolution into today's massive clusters \citep{Arnaud09}.

Entropy evolution from the formation epoch onwards is the key to disentangling the various non-gravitational processes: cooling and heating via SMBH feedback and supernova-driven galactic winds. IXO will measure the gas entropy and metallicity of clusters to $z \sim 2$\ to reveal whether the excess energy observed in present-day clusters was introduced early in the formation of the first halos or gradually over time, crucial input to our understanding of galaxy and star formation. 

Measuring the evolution of the metal content and abundance pattern of the IGM with IXO will show when and how the metals are produced, in particular the relative contribution of Type Ia and core-collapse supernovae, and the stellar sources of carbon and nitrogen. Precise abundance profiles from IXO measurements will constrain how the metals produced in the galaxies are ejected and redistributed into the intra-cluster medium. 

%\paragraph
{\it Cosmology}\ 
The mystery of Dark Energy can be studied by observing the expansion history of the Universe and the growth of matter density perturbations. X-ray observations of galaxy clusters with IXO will provide both tests, complementing other planned cosmological experiments \citep{Vikhlinin09}. Combining the distance-redshift relation $d(z)$\ and growth of structure data will dramatically improve constraints on the Dark Energy equation of state, comparable to the constraints from other stage IV Dark Energy experiments planned for this timeframe. These IXO data also test whether the cosmic acceleration is caused by modifications to Einstein's theory of gravity on large scales.

Galaxy cluster observations also provide their own $d(z)$\ test by assuming that the mass fraction of hot intracluster gas is constant with redshift. IXO will provide the precise temperature measurements essential to determine the cluster masses. IXO observations of 500 relaxed clusters will give an independent $d(z)$\ measurement. The spectral capabilities of IXO will provide direct checks on the relaxed state of the cluster through velocity measurements of the intra-cluster medium.

%A recent advance in using galaxy clusters for cosmology was made by combining Chandra %observations with advances in numerical modeling, leading to new dark energy constraints from both %geometric and growth of structure methods. Similarly, combining weak lensing and IXO observations %of high-z clusters will reduce systematic errors sufficient to constrain the growth factor to 1\% accuracy %throughout z = 0-2, leading to very competitive uncertainties in cluster-based cosmological %measurements. 

%\paragraph
{\it The Cosmic Web of Baryons}\ 
Less than 10\% of the baryons in the local Universe lie in galaxies as stars or cold gas, with the remainder predicted to exist as a dilute gaseous filamentary network, ``the cosmic web.''  Some of this cosmic web is detected in Ly$\alpha$\ and O VI absorption lines, but half remains undetected. Growth of structure simulations predict that these "missing" baryons are shock heated up to temperatures of $10^7$\,K in unvirialized cosmic filaments and chemically enriched by galactic superwinds. The order of magnitude increase in collecting area and R = 3000 spectral resolution of IXO is required to enable detection of the missing baryons and characterize their velocity distribution along at least 30 lines of sight \citep{Bregman09}. This distribution of mass as a function of temperature can be determined from X-ray absorption line grating spectroscopy of highly ionized C, N, and O detected against background AGNs. %The extent and nature of galactic superwinds that enrich the web will also be measured both from the proximity of absorption sites to galaxies and the dynamics of the hot gas.

%Most galaxies have lost more than 2/3 of their baryons, relative to the cosmological ratio of baryons to dark matter. These missing baryons are probably hot, but we do not know if they were expelled as part of a starburst-phase galactic wind, or pre-heated so that they simply never coalesced. X-ray absorption line observations with IXO will, for the first time, identify the location and metallicity of these Local Group baryons from the line centroids and equivalent widths of hot C, N, and O ions, revealing a crucial aspect of galaxy formation (Bregman et al.).

\begin{figure}
\includegraphics[totalheight=2in]{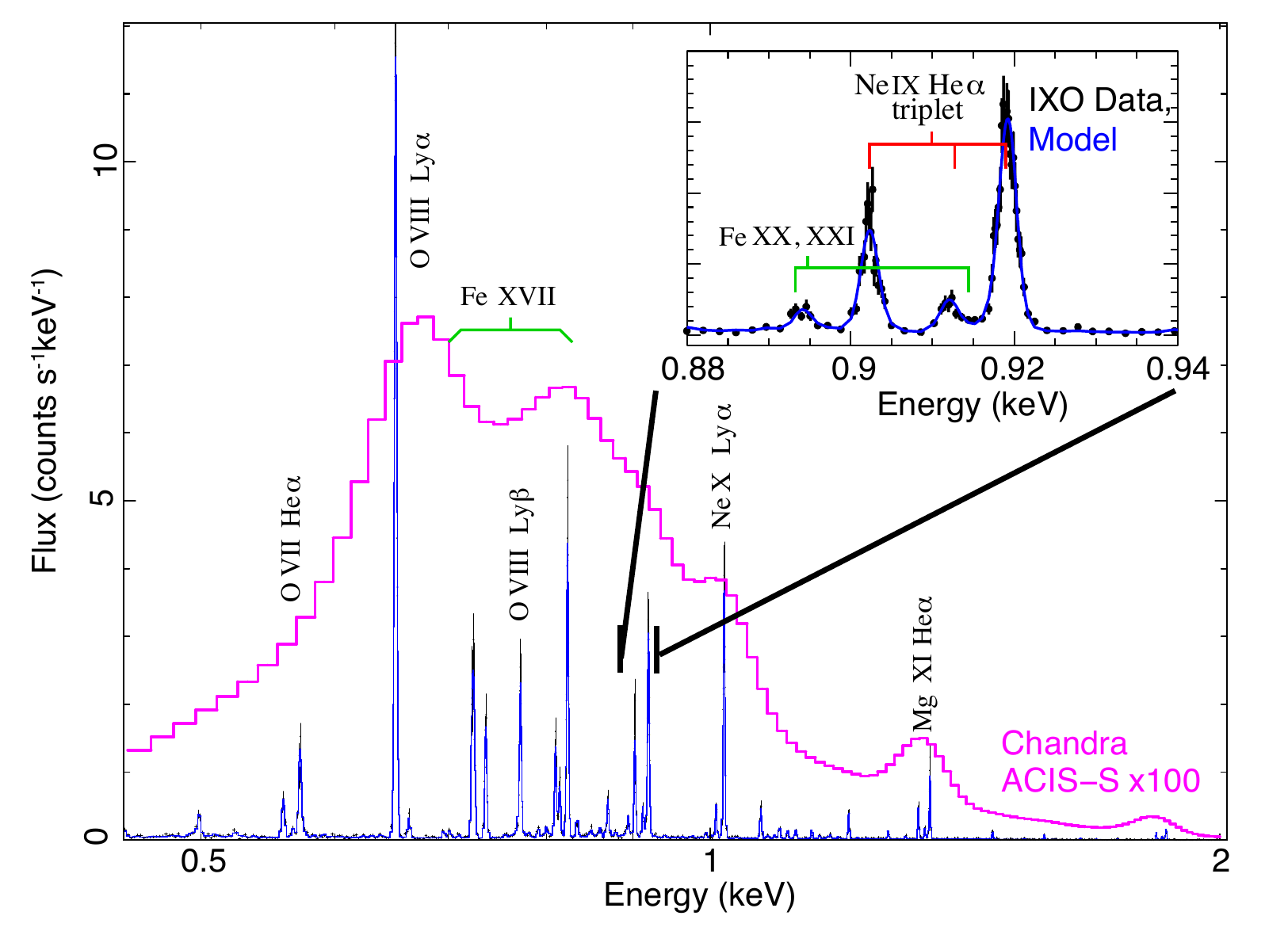}
\includegraphics[totalheight=2in]{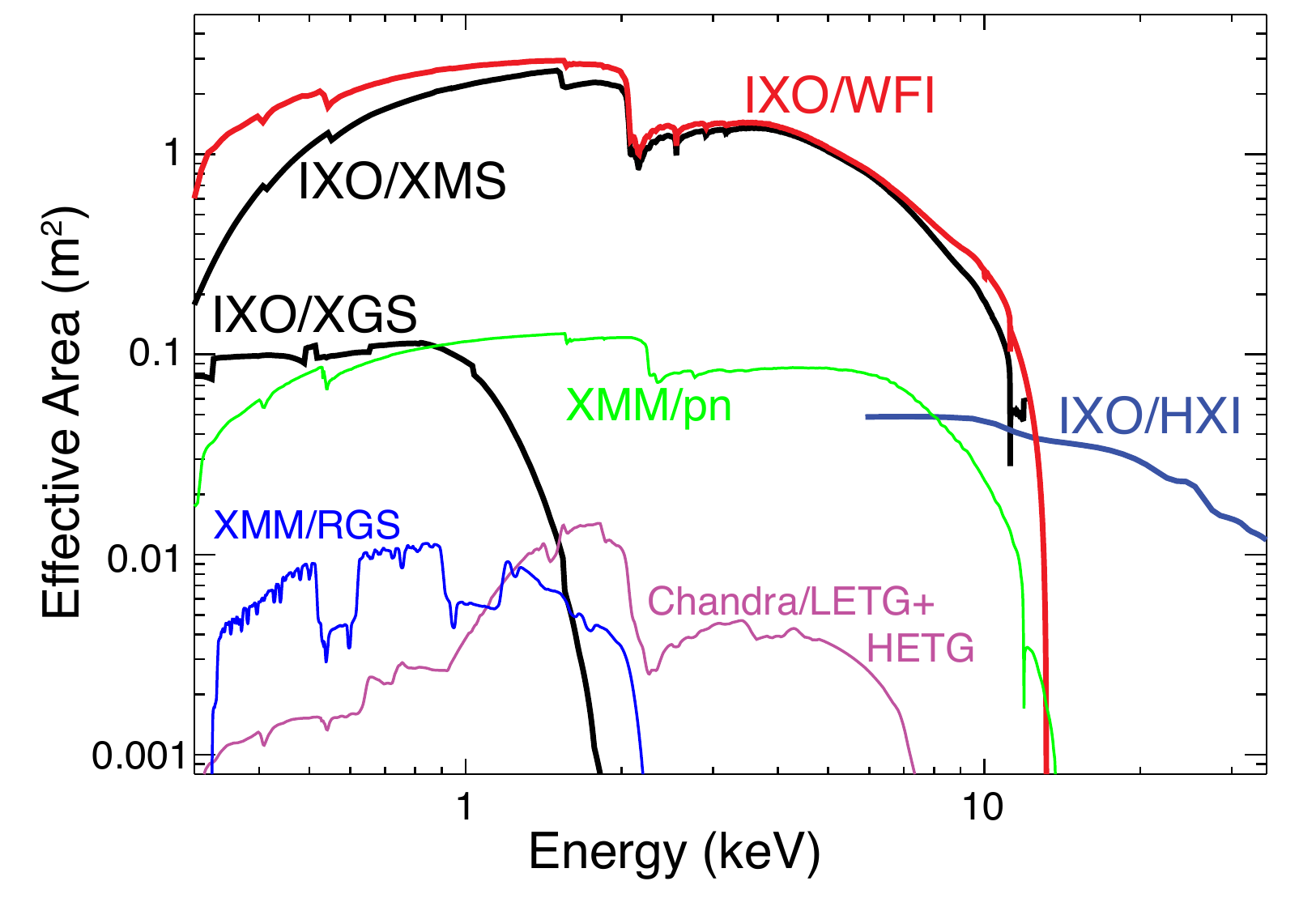}
\caption{[Left] The IXO XMS spectrum of a 1 sq arcmin region within a typical nearby starburst superwind, showing the relative strength of the line to continuum emission.  [Right] The IXO effective area will be more than an order of magnitude greater than current imaging X-ray missions. Coupling this with an increase in spectral resolving power up to two orders of magnitude higher relative to previous capabilities will open a vast discovery space for high-energy phenomena.}\label{fig:SBandEA}
\end{figure}

%\paragraph
{\it Life Cycles of Matter and Energy}\ 
The dispersal of metals from galaxies can occur as starbursts drive out hot gas that is both heated and enriched by supernovae. IXO is needed to measure the hot gas flow velocity using high-throughput spectroscopic imaging (Fig.~\ref{fig:SBandEA}[Left]), and in turn determine the galactic wind properties and their effects \citep{Strickland09}. The distribution of metal abundances in the Milky Way, including both the gas and dust components, will be mapped using absorption line measurements along hundreds of lines of sight \citep{Lee09}. On smaller scales, emission from gaseous remnants of supernovae seen with IXO will offer a comprehensive three-dimensional view of the ejecta composition and velocity structure, allowing detailed studies of nucleosynthesis models for individual explosions \citep{Hughes09}. 

IXO will reveal the influence of stars on their local environment via measurements of their coronal activity and stellar winds \citep{Osten09}. This influence also includes their effect on habitable zones as well as on planet formation. Observations of star-forming regions have shown that X-rays from stellar flares irradiate protoplanetary disks, changing the ion-molecular chemistry as well as inducing disk turbulence. While Chandra has detected a few immense flares, the most significant impact on the protoplanetary disk is in the integrated output of smaller flares, which can only be characterized by IXO \citep{Feigelson09}.

\section{Mission Implementation}

%\begin{figure}
%\includegraphics[totalheight=1.5in]{Spacecraft.pdf}
%\includegraphics[totalheight=2in]{IXO_Articulated_Deployed_003.pdf}
%\caption{Artists' conceptions of the US [Left] and ESA [Right] design concepts for IXO\label{fig:spacecraft}}
%\end{figure}

As part of submitting IXO to the ESA Cosmic Visions and the US Astro2010 processes, both NASA and ESA have developed detailed spacecraft concepts that are very similar.
%NASA and ESA have each developed a detailed spacecraft concept 
%(Fig.~\ref{fig:spacecraft}) 
%that is very similar. 
IXO is built around a large area grazing-incidence mirror assembly with a 20 m focal length. Flight-proven extending masts deploying the focal plane allow the observatory to fit into the launch vehicle fairing (an Atlas V or Ariane V). The IXO payload consists of 1) the Flight Mirror Assembly (FMA), a large area grazing incidence mirror; 2) four instruments mounted in the mirror focal plane on a movable instrument platform (MIP), which are placed in the mirror focus one at a time; and 3) an X-ray Grating Spectrometer (XGS) that intercepts and disperses a fraction of the beam from the mirror onto a CCD camera, operating simultaneously with the observing MIP instrument \citep{Bookbinder10}. An L2 orbit facilitates high observational efficiency and provides a stable thermal environment. The mission design life is five years, with consumables sized for 10 years. Both studies 
found %concluded 
that the IXO spacecraft could be built with technologies that are fully mature today.

The 
%Flight Mirror Assembly 
FMA provides effective area of 3 m$^2$\ at 1.25 keV, 0.65 m$^2$\ at 6 keV, and 150 cm$^2$\ at 30 keV. The effective area is compared to past observatories in Fig.~\ref{fig:SBandEA}[Right]. To meet the 5 arc sec mission-level half-power diameter (HPD) requirement for the observatory across the field of view, the FMA angular resolution must be 4 arc sec or better. Attaining the large effective area within the launch vehicle mass constraint requires a mirror with a high area-to-mass ratio: 20 cm$^2$/kg, 50 times larger than Chandra and eight times larger than XMM-Newton. Two possible mirror  technologies are being developed in a coordinated fashion by NASA, ESA and JAXA as a risk reduction strategy. These are thermally formed (``slumped'') glass and silicon pore optics. Both approaches lead to a highly modular mirror design, where the key technology hurdle is the construction of a module. 
%The observatory can accommodate either mirror approach. 
Both technologies have demonstrated X-ray performance of $\sim 15$\,arcsec HPD, with plans underway to demonstrate the required performance.

The X-ray Microcalorimeter Spectrometer (XMS) provides high spectral resolution, non-dispersive imaging spectroscopy over a broad energy range. The driving performance requirements are to provide spectral resolution of 2.5 eV over the central $2\times 2$\, arcmin in the 0.3-7.0 keV band, and 10 eV to the edge of the $5\times 5$ arcmin field of view. The XMS is composed of an array of microcalorimeters, devices that convert individual incident X-ray photons into heat pulses and measure their energy via precise thermometry. The microcalorimeters are based on Transition-Edge Sensor (TES) thermometers. Currently, 2.3 eV spectral resolution has been demonstrated in a non-multiplexed TES and 2.9 eV has been achieved in a $2\times 8$ array using a time-division SQUID multiplexer system. A Continuous Adiabatic Demagnetization Refrigerator (CADR) and a mechanical cryocooler provide cooling to 50 mK without expendable cryogens. 

The Wide Field \& Hard X-ray Imager (WFI/HXI) are two detectors incorporated into one instrument, with the HXI mounted directly behind the WFI. The WFI is an imaging X-ray spectrometer with an $18\times 18$\ arcmin field of view. It provides images and spectra in the 0.1-15 keV band, with nearly Fano-limited energy resolution (50 eV at 300 eV,$ < 150$\,eV at 5.9 keV). The 1 arcsec pixel size oversamples the beam and thus minimizes pulse pile up. The WFI's key component is the DEPFET (Depleted P-channel Field Effect Transistor) Active Pixel Sensor (APS). Compared with earlier CCD-type detectors, the APS concept has the significant advantage that the charge produced by an incident X-ray photon is stored in and read directly from each pixel, which reduces readout noise, and offers radiation hardness. Prototype DEPFET devices of $64\times 64$ pixels have been tested successfully; an energy resolution at 5.9 keV of 126 eV has been demonstrated. The HXI extends IXO's energy coverage to 40 keV with an energy resolution $\sim$ 1 keV (FWHM) at 30 keV and a FOV of $12\times 12$\ arcmin. The HXI is a $7\times 7$\ cm wide Double-sided Strip Cadmium Telluride (DS-CdTe) detector, based on those to be flown on ASTRO-H.

The X-ray Grating Spectrometer (XGS) is a wavelength-dispersive spectrometer for high-resolution spectroscopy, offering spectral resolution ($\lambda/\Delta \lambda$) of 3000 (FWHM) and effective area of 1000 cm$^2$\ across the 0.3-1.0 keV band. The arrays of gratings intercept a portion of the converging FMA beam and disperse the X-rays onto a CCD array. Two viable grating technologies reduces risk. One utilizes Critical Angle Transmission (CAT) gratings with heritage from the Chandra High Energy Transmission Grating, but substantially higher efficiency. Another approach uses ``off plane'' reflection gratings based on XMM-Newton gratings. To give higher performance the IXO gratings are ruled along the direction of incidence rather than perpendicular to it as 
%was implemented 
on XMM-Newton.

The High Time Resolution Spectrometer (HTRS) will perform precise timing measurements of bright X-ray sources with fluxes of $10^6$\ counts per second in the 0.3-10 keV band, while providing moderate spectral resolution (200 eV FWHM at 6 keV). The X-ray Polarimeter (XPOL) is an imaging polarimeter, with polarization sensitivity of 1\% for a source with 2-6 keV flux of $5\times10^{-12}$\, ergs cm$^{-2}$s$^{-1}$ (1mCrab). XPOL utilizes a fine grid Gas Pixel Detector to image the tracks of photoelectrons produced by incident X-rays%and determine the direction of the primary photoelectron
, which convey information about the polarization.

%%%%%%%%%%%%%%%%%%%%%%%%%%%%%%%%%%%%%%%%%%%%
%% Sample figure:
%%
%% The option [height=...] scales the picture to the given height,
%% without it it would be printed at its nominal size
%%%%%%%%%%%%%%%%%%%%%%%%%%%%%%%%%%%%%%%%%%%%

%\begin{figure}
 % \includegraphics[height=.3\textheight]{golfer}
  %\caption{Picture to fixed height}
%\end{figure}

%%%%%%%%%%%%%%%%%%%%%%%%%%%%%%%%%%%%%%%%%%%%%%%%
%% BACKMATTER
%%%%%%%%%%%%%%%%%%%%%%%%%%%%%%%%%%%%%%%%%%%%%%%%

\begin{theacknowledgments}
This document could not have been written without the tireless efforts of the IXO Science Coordination Group, Science Definition Team, Telescope Working Group, Instrument Working Group, and the contributions of many other IXO science associates.
\end{theacknowledgments}



\bibliographystyle{aipprocl} % if natbib is missing

%%%%%%%%%%%%%%%%%%%%%%%%%%%%%%%%%%%%%%%%%%%
%% The following lines show an example how to produce a bibliography
%% without the help of the BibTeX program. This could be used instead
%% of the above.
%%%%%%%%%%%%%%%%%%%%%%%%%%%%%%%%%%%%%%%%%%%

\end{document}